---------------------------Latex
file-------------------------------------------

\documentstyle[equations,12pt]{article}
\newcommand{\ba}{\begin{array}}
\newcommand{\ea}{\end{array}}
\newcommand{\bc}{\begin{center}}
\newcommand{\ec}{\end{center}}
\newcommand{\be}{\begin{equation}}
\newcommand{\ee}{\end{equation}}
\newcommand{\plb}[1]{Phys. Lett. {\bf #1}}

\newcommand{\npb}[1]{Nucl. Phys. {\bf #1}}

\newcommand{\cH}{{\cal H}}

\def\bild#1\over#2{\mathrel{\mathop{\kern0pt #1}\limits_{#2}}}
\renewcommand{\theequation}{\arabic{section}.\arabic{equation}}
\begin{document}

\begin{flushright}
IPNO/TH 94-01
\end{flushright}
\vspace{1 cm}
\begin{center}
{\large {\bf $SU_3$ coherent state operators
and invariant correlation \\
functions and their quantum group counterparts}}
\vspace{1 cm}

Hagop Sazdjian  \\
{\it Division de Physique Th\'eorique\footnote[1]{Unit\'e
 de Recherche des Universit\'es Paris 11 et Paris 6 associ\'ee au CNRS},
         Institut de Physique Nucl\'eaire, \\
Universit\'e Paris XI,\\
F-91406 Orsay Cedex, France}\\
\vspace{1 cm}
Yassen S. Stanev\footnote[2]{On leave of absence from the Institute for
Nuclear Research and Nuclear Energy, Tsarigradsko Chauss\'ee 72, BG-1784
Sofia, Bulgaria}\\
{\it Laboratorio Interdisciplinare per le Scienze Naturali ed Umanistiche,\\
Scuola Internazionale Superiore di Studi Avanzatti, \\
I-34014 Trieste, Italy} \\
\vspace{1 cm}
Ivan T. Todorov$^2$\\
{\it Laboratoire de Physique Th\'eorique et Hautes
Energies\footnote[3]{Laboratoire associ\'e au CNRS},\\
  Universit\'e Paris XI,\\
F-91405 Orsay Cedex, France}

\end{center}
\vspace{5 cm}

\noindent
{ }
\newpage

\begin{center}
{\large Abstract}
\end{center}
Coherent state operators $(CSO)$ are defined as operator valued functions on
$G =SL(n,C)$ being homogeneous
with respect to right multiplication by lower triangular matrices. They act
on a model space containing all
 holomorphic finite dimensional representations of $G$ with
 multiplicity 1. $CSO$ provide an analytic tool for studying $G$ invaraiant 2-
and 3-point functions,
 which are written down in the case of $SU_3$. The quantum group deformation
 of the construction gives rise to a non-commutative coset space.
We introduce a ``standard'' polynomial basis in this
 space (related to but not identical with the Lusztig  canonical basis)
which is appropriate for writing down $U_q(s\ell _3)$ invariant 2-point
functions for representations of the type $(\lambda ,0)$ and $(0,\lambda )$.
General invariant 2-point functions are written down in a mixed
Poncar\'e-Birkhoff-Witt type basis.\par

\newpage

\tableofcontents

\newpage

 \section{Introduction}
 Extending the applications of the quantum universal enveloping algebra
$U_q(s\ell_2)$ to rational
 conformal models  with braid group statistics \cite{aggs,ps,gpfgp,mr,tghpt,fk}
to the higher rank case encounters two
 types of difficulties. At  the classical (undeformed) level the n-point
invariants for higher rank groups are not
 known  and appear hard to classify. At the q-deformed level we have to deal
in addition with a non-commutative  coset space.

A powerful analytic tool for writing down (classical and quantum) n-point
invariants
 and $U_q$ exchange relations is provided by the notion of a coherent
 state operator (CSO). This is the underlying concept behind the polynomial
realization of finite
 dimensional representations (used in this context in Refs. \cite{zf,gpfgp}).
It was spelled out and applied to the
 simplest rank  1 (quantum)  algebra in Refs. \cite{fst,sth}.
Here we extend  this construction to the case
 of $SU_{r+1}$ and $U_q(s \ell_{r+1})$ ($r$ being the rank).

The next step, computing n-point invariants, reveals a complexity that
increases with the rank.
 Therefore, we restrict attention to the rank 2 case, $SU_3$ and
\be \label{1e1}
U_q \equiv U_q (s\ell_3)\ .\ee

In order to define $U_q$ CSO we need the dual object, the quantum
 group $SL_q(3)$. More generally, $SL_q(n)$ is defined as a Hopf algebra with
$n^2$  generators subject to $\left (^n_2\right )$ commutation relations
(CR) and an n-linear normalization
 condition stating that the quantum determinant of the matrix
($T^\alpha _\beta$) is one. The CR are
 expressed in terms of the $U_q(s\ell_n)$ R-matrix computed for the
coproduct of the defining
(n-dimensional) representation of the quantum algebra -see Ref. \cite{frt}.
The next step, the identification of a
 quantum Borel subgroup $B_q$ and of the corresponding $q$-coset space
$SL_q(n) /B_q$, has
 also  been  sketched for  an arbitrary $n$ -see Ref. \cite{it}.
In order  to avoid unnecessary complications and bulky  formulae we restrict
from the outset  attention to the case $n=3$.

We start  in Sec. 2 by introducing our version of coherent states (CS).
We write down a closed form polynomial
 expression for the general $SU_3$ invariant 3-point function (Proposition
2.1).

The central object of this paper, the $U_q$ homeogeneous space
\be \label{1e2}
N_q=SL_q(3)/B_q\ ,\ee
is introduced and studied in Sec. 3. Quantum CSO are defined in Sec. 4,
where we write
down $U_q$ invariant 2-point functions introducing on the way
a new standard basis, which is a variant of the Lusztig canonical
basis \cite{l}.

\newpage

\section{Classical coherent states}
\setcounter{equation}{1}
The  concept of a coherent state, originally associated with a nilpotent
Lie group,
goes back to Hermann Weyl and has been the subject of a long evolution
(see Ref. \cite{p} and
 references therein). The most familiar example is given by the Fock space
vector \[\Phi(\zeta)=e^{\zeta a^*}|0> \]
where $a^*$ is a creation operator, its adjoint annihilating the Fock
vacuum : $a|0>=0$.
 A characteristic property for such a vector valued function is provided by
the equation \[ (a^* - \frac{d}{d\zeta}) \Phi (\zeta) =0\ .\]
In the  case of a simple compact Lie group G the role of $a^*$ is played by
the Cartan-Weyl
raising operators $E_\alpha$ of the complexified Lie algebra. For an
irreducible finite dimensional
 representation of $G$, the counterpart of $\Phi$ appears as a polynomial
function of its  arguments.

\subsection{CSO for $SU_3$. Transformation properties}

To fix the ideas we begin by considering the induced representation of
the complexification
$SL(3,C)$ of the simple  compact Lie group $SU_3$ with a Borel inducing
subgroup B of lower
 triangular matrices. The homogeneous  space $SL(3,C)/B$ has a dense
open set isomorphic to the subgroup $N$ of upper
 triangular matrices with units on the diagonal :
\setcounter{equation}{0}
\be \label{2e1}
N= \left\{ Z= \left( \ba{ccc} 1& \zeta_1& \zeta_{12} \\
0&1&\zeta_2 \\ 0&0&1\ea \right) \right\}\ . \ee
For each highest weight $\underline{\lambda}=
(\lambda_1, \lambda_2) $ ($\lambda_i=0,1,\ldots)$
 of $SU_3$ we define a CSO $\Lambda=\Lambda(g)$ as follows. Let
$\chi_{\underline{\lambda} }$ be a character (i.e., a 1 dimensional
representation)
 of the inducing subgroup B given by \be \label{2e2}
\chi_{\underline{\lambda} }(b)= \beta_1^{-\lambda_2} \beta_2^{-\lambda_1}
\quad {\rm for} \quad
b = \left(\ba{ccc}  \beta_1&0&0 \\
b_1& \beta_1^{-1} \beta_2& 0 \\ b_{12}& b_2& \beta^{-1}_2 \ea \right)\ .\ee
$\Lambda=\Lambda(g)$ is a function on $SU_3$ with values in an algebra of
operators on a Hilbert space
 (to be introduced in Subsection 2), satisfying the homogeneity condition
\be \label{2e3}
\Lambda (gb)= \Lambda (g) \chi_{\underline{\lambda}} (b)\ .\ee
It follows that $\Lambda$ is determined by its values on the subgroup $N$,
(\ref{2e1}). Using  Eqs.(\ref{2e2})
 and (\ref{2e3}) and the Gauss decomposition
\be \label{2e4}
gZ= Z_g b(g, \zeta )\ ,\ \ Z_g \in N\ ,\ \ b(g, \zeta) \in B \ee
(which assumes $g_{11} \neq 0$), we deduce the following  transformation law
 for $\Lambda (\zeta_1, \zeta_{12}, \zeta_2)$ under the various subgroups of
$ SL(3, C)$ :
\subequations
\be \label{2e5a}
Z':\ \ \Lambda(\zeta_1,\zeta_{12}, \zeta_{2})\longrightarrow \Lambda(\zeta_1+
\zeta_1',
\zeta_{12}+ \zeta_1'\zeta_2+ \zeta_{12}',\zeta_2+ \zeta_2' ) \; ,\ee

\be \label{2e5b}
h_\alpha =\left( \ba{ccc} \alpha_1&0&0\\ 0&\alpha_1^{-1}\alpha_2&0 \\ 0&0&
\alpha_2^{-1} \ea \right):\ \
\Lambda(\zeta_1, \zeta_{12}, \zeta_2) \to \Lambda(\alpha^2_1 \alpha_2^{-1}
\zeta_1, \alpha_1
 \alpha_2 \zeta_{12} ,
\alpha_1^{-1} \alpha_2^2 \zeta_2)\ \alpha_1^{-\lambda_2} \alpha_2^{-\lambda_1}
\ ,\ee

\[ b=\left(  \ba{ccc} 1&0&0\\ b_1&1&0 \\ 0&b_2&1 \ea \right) :\ \
\Lambda(\zeta_1, \zeta_{12}, \zeta_2) \to \Lambda(\tilde \zeta_1 ,
\tilde \zeta_{12},
\tilde \zeta_2)\ \beta_1^{ -\lambda_2} \beta_2^{-\lambda_1} \;\; ,\]

\be \label{2e5c} \ba{l}
\tilde \zeta_1= \beta_1\{ \zeta_1+b_2(\zeta_1\zeta_2- \zeta_{12})\} \; ,\ \
\tilde\zeta_2=\beta_2(\zeta_2+ b_1\zeta_{12}) \; , \\
  \\
\tilde \zeta_{12}= \beta_{2}\zeta_{12} \; ; \ \ \beta^{-1}_1=1+ b_1 \zeta_1+
b_1b_2(\zeta_1\zeta_2-\zeta_{12})\ ,\ \  \beta^{-1}_2 =1+b_2 \zeta_2 \; .
\ea \ee
\endsubequations

Denoting the $s\ell_3$ Chevalley generators corresponding to the
 paramaters $\zeta'_1, \zeta'_2$, $\ln \alpha_1$ , $ \ln \alpha_2$, $b_1,
b_2$ by
$E_1,E_2,H_1, H_2, F_1 ,F_2$  we can write the associated infinitesimal law :

\subequations
\be \label{2e6a}
[E_1, \Lambda] = (\partial_1+ \zeta_2 \partial_{12}) \Lambda\ ,\ \
[E_2, \Lambda] = \partial_2 \Lambda \ \
 (\partial_i\equiv \frac {\partial}{\partial \zeta_i}) \; ; \ee

\[
[ H_1, \Lambda ]= (2 \zeta _1\partial _1+ \zeta _{12} \partial _{12}-
\zeta _2 \partial _2 -\lambda _2) \Lambda \; ,\]
\be \label{2e6b}
[ H_2, \Lambda ]= (2 \zeta _2\partial _2+ \zeta _{12} \partial _{12}-
\zeta _1 \partial _1 -\lambda _1) \Lambda \; ;
\ee

\[
[F_1, \Lambda] = (\lambda_2\zeta_1-\zeta^2_1 \partial_1+\zeta_{12}
\partial_2) \Lambda \; , \]
\be \label{2e6c}
[F_2, \Lambda] = (\lambda_1\zeta_2+( \zeta_1 \zeta_2 -\zeta_{12})
\partial_1 - \zeta^2_2 \partial_2
-\zeta_2 \zeta_{12} \partial_{12}) \Lambda \; .
\ee

The action of the two remaining infinitesimal operators is derived from
their definitions:
\be \label{2e6d} \ba{lll}
E_{12} &=& [E_1,E_2] \Rightarrow [E_{12}, \Lambda]= [\partial_2, \partial_1+
\zeta_2 \partial_{12} ]\Lambda = \partial_{12} \Lambda \; , \\
F_{12} &=& [F_2,F_1] \Rightarrow [F_{12}, \Lambda]= [ \zeta_{12}
(\lambda_1- \zeta_2 \partial_2)+ (\zeta _1 \zeta _2 -\zeta _{12})
(\zeta_1\partial_1 -\lambda_2) - \zeta^2_{12} \partial_{12}] \Lambda \; .\ea
\ee
\endsubequations

We  recall that the Chevalley generators satisfy the commutation relations
(CR) :
\subequations
\be \label{2e7a}
[H_i,E_j]=c_{ij} E_j\ ,\ \  [H_i , F_j] =-c_{ij} F_j \; ,
\ee
\be \label{2e7b}
[E_i,F_j]=\delta_{ij} H_i\ ,\ \  [E_i , E_{12}] =0\ ,\ \
[F_i, F_{12}]=0  \; ,
\ee
\endsubequations
where $(c_{ij}$) is the Cartan matrix

\be \label{2e8}
(c_{ij})= \left( \ba{cc} 2&-1\\ -1& 2\ea \right)\ . \ee

\subsection{A model Hilbert space. Coherent state vectors. 2-point functions}

We now proceed to describing an operator realization of $\Lambda(\zeta)$.

Let $\cH$ be a model Hilbert space for the finite dimensional
(unitary) irreducible representations (IR) of $SU_3$ :
\be \label{2e9}
\cH =\oplus_{\underline {\lambda}} \cH_{\underline {\lambda}} \; ,
\ \ {\rm dim} \cH_{\underline {\lambda}} =
\frac{1}{2} (\lambda_1+1)(\lambda_2+1) (\lambda_1+\lambda_2+2) \ , \ee
each $\underline {\lambda}$ entering with multiplicity 1.
Let $U(g)$ be the corresponding (infinite dimensional)
 fully reducible representation of $SU_3$. The covariance  law (2.5)
for CSO can be summed up by
\be \label{2e10}
U(g)\Lambda(Z) U(g)^{-1} = \Lambda(gZ)=
\Lambda(Zg) \chi_{\underline{\lambda}} (b(g, \zeta))\; , \ee
where $Z_g$ and $b(g,\zeta)$ are defined by the Gauss decomposition
(\ref{2e4}) (and computed for the various subgroups of $SL(3, C)$ in
(2.5)).

We single out a unit vector $|0>$, the $SU_3$ vacuum, in the 1-dimensional
$SL(3,C)$ invariant
 subspace ${\cH}_{\underline {0}}$ of $\cH$. The action of a CSO on $|0>$
is determined -up to normalization- by its transformation properties.

We recall the Gel'fand-Cetlin pattern notation \cite{gcbr} for the $SU_3
\supset U_2 \supset U_1$ chain :
\be \label{2e11} \left(
\ba{ccccc} \lambda_1+\lambda_2 &&\lambda_2 && 0\\
 &m_1&&m_2& \\
 &&m_3&&\ea \right) \quad \ba{l} 0\leq m_2 \leq \lambda_2 \\
\lambda_2 \leq m_1 \leq \lambda_1+\lambda_2 \\
m_2 \leq m_3 \leq m_1 \ea \ .
\ee
Such a pattern represents an eigenvector of $H_1$ and $H_2$ with eigenvalues
$2m_3-m_1-m_2$  and
$2m_1+2m_2-m_3-\lambda_1-2\lambda_2$, respectively. The CSO $\Lambda(Z)
\equiv \Lambda (\zeta)$  will be  normalized by the condition
\subequations
\be \label{2e12a}
\Lambda (\zeta =0) |0> = |-\lambda_2, -\lambda_1> \quad , \ee
where the right hand side is a shorthand for the (unit) lowest weight vector
\be \label{2e12b}
|-\lambda_2, -\lambda_1>= \left( \ba{ccccc} \lambda_1+\lambda_2&&\lambda_2&&0
\\
 &\lambda_2& & 0&\\ &&0&&\ea \right) \ee
characterized by
\be \label{2e12c} \ba{r}
(H_1+\lambda_2)|-\lambda_2,-\lambda_1>=0=(H_2+\lambda_1)|-\lambda_2,
-\lambda_1>
=F_i|-\lambda_2,-\lambda_1> \\
(i=1,2) \ea \ .\ee
\endsubequations
Under these conditions one can derive the expression for the $\zeta$ dependent
 coherent state
\be \label{2e13}
\Lambda(\zeta)|0>\ =\ e(\zeta_2 E_2)\ e(\zeta_{12} E_{12})\ e(\zeta_1E_1)\
\vert -\lambda_2,-\lambda_1> \ , \ee
where $e(X)$ is the exponential function.

Indeed, applying $E_i$ to both sides and using the definition (2.6d) of
 $E_{12}$ and the commutation  property (2.7b), it is straightforward to
recover
 the infinitesimal transformation law (2.6a). The remaining $s\ell_3$
transformation properties of the vector (\ref{2e13}) are verified
in a similar manner. These properties, together with the ``initial
 condition'' (2.12), fix completely the CS vector.

The CS (\ref{2e13}) is actually a polynomial in $(\zeta_\alpha)$ which
can be written in the following factorized form :
\begin{eqnarray} \label{2e14}
\Lambda (\zeta)|0> &=& \sum_{m_2+n_2\leq \lambda_2} \zeta^{m_2}_2
\zeta^{n_2}_{12} \frac{E_2^{m_2}}{m_2!} \frac{E_{12}^{n_2}}{n_2!}
|-\lambda_2, 0> \otimes \nonumber \\
  &\ &\ \  \sum_{m_1+n_1\leq \lambda_1} \zeta^{m_1}_1
(\zeta_{12}-\zeta_1\zeta_2)^{n_1}
\frac{E_1^{m_1}}{m_1!} \frac{E_{12}^{n_1}}{n_1!} |0,-\lambda_1>\ ;
\end{eqnarray}
here we have used the identity
\be \label{2e15}
e(\zeta_2 E_2)\ e(\zeta_{12} E_{12})\ e(\zeta_1 E_1)\ =\ e(\zeta_1E_1)\
e((\zeta_{12}-\zeta_1\zeta_2)E_{12})\ e(\zeta_2E_2) \quad . \ee

The highest weigth vector entering the expansion (\ref{2e14}) is the
coefficient to
  $\zeta^{\lambda_2}_{12} (\zeta_{12}- \zeta_1\zeta_2)^{\lambda_1} $ :
\subequations
\be \label{2e16a}
|0, \lambda_2> = \frac{E_{12}^{\lambda_2}}{\lambda_2!} |- \lambda_2, 0>\ ,\ \
|\lambda_1, 0>
= \frac{E^{\lambda_1}_{12} }{\lambda_1!} |0,-\lambda_1> \quad , \ee
\be \label{2e16b}
|\lambda_1, \lambda_2> = |\lambda_1,0> \otimes |0,\lambda_2>=
\left( \ba{lllll}\lambda_1+\lambda_2& &\lambda_2& &0\\
 &\lambda_1+\lambda_2& &  \lambda _2& \\
 & &\lambda_1 + \lambda_2& & \\
\ea \right) \ , \ee
\be \label{2e16c}
(H_1-\lambda_1) |\lambda_1,\lambda_2> =(H_2- \lambda_2) |\lambda_1,
\lambda_2> =0=E_\alpha
|\lambda_1, \lambda_2> \ . \ee
\endsubequations
We shall also use the bra CS vector
\be \label{2e17}
<0| \Lambda(\zeta) = < \lambda_2\lambda_1|e(-\zeta_1 E_1)e (-\zeta_{12}
E_{12}) e(-\zeta_2E_2) \ . \ee
We find it convenient in what follows to use the notation $\Lambda$
for both the CSO and the associated representation of $SU_3$
(interchanging in the second meaning $\Lambda $ with $\underline{\lambda}$).

The inner product of two CS vectors corresponding to the representations
$\Lambda$ and
 $\Lambda'$ is only nonzero if $\Lambda'$ is the conjugate representation to
$\Lambda$ : $\Lambda'=(\lambda_2,\lambda_1)\ (\equiv \bar \Lambda) $; indeed,
\subequations
\be \label{2e18a}
<0|\Lambda '(Z')\Lambda (Z)|0>\ =\ \left [P(\zeta ',\zeta )\right ]^
{\lambda _1} \left [-P(\zeta ,\zeta ')\right ]^{\lambda _2}
\delta _{\lambda _1 \lambda _2 '} \delta _{\lambda _2 \lambda _1 '}\ ,
\ee
where \be \label{2e18b}
P(\zeta ',\zeta) \ =\ \zeta _{12}-\zeta _{12}'-\zeta _1' (\zeta _2-\zeta _2')\
=\ \zeta _{12}+\zeta _{21}'-\zeta _1'\zeta _2 \ee
with
\be \label{2e18c}
\zeta _{21}\ =\ \zeta _1\zeta _2 - \zeta _{12}\ .
\ee
\endsubequations

To define $\Lambda$ on a vector belonging to a nontrivial IR
$\Lambda^{(i)}$, one has
 to specify a nonvanishing matrix element of the type
\be \label{2e19}
<\lambda_1^{(f)}, \lambda_2^{(f)}| \Lambda(\zeta)|- \lambda_2^{(i)},
-\lambda_1^{(i)}> \ . \ee
(In the case of $SU_2$ it is enough to specify the weight
$\lambda^{(f)} =2I^{(f)}$ of the
 final state, the counterpart of (\ref{2e19}) being then a multiple of
$\zeta^{I^{(i)}+I-I^{(f)}}$.
 On the other hand, one knows -see Subsection 3 below- that there could  be
two linearly
independent $SU_3$ invariant 3-point functions for a given triplet  of $SU_3$
weights.)
The problem is  thus to classify all 3-point invariants of $SU_3$ (and then,
 hopefully, to get a hold on the $n$-point invariants).

\subsection{$SU_3$ invariant 3-point functions}
The general $SU_3$ invariant 2-point function is proportional to
the right-hand side of Eq. (2.18a). Using invariance under $Z$ shifts and the
relation
\[
\ba{ll}
Z^{-1} Z' & =
\left(
\ba{ccc} 1&-\zeta_1& \zeta_1\zeta_2-\zeta_{12} \\ 0&1&-\zeta_2 \\ 0&0&1 \ea
\right)
\left(
\ba{ccc} 1&\zeta_1'&\zeta_{12}' \\ 0&1&\zeta_2' \\ 0&0&1 \ea\right) \\
 & \\
&= \left(
\ba{ccc} 1&\zeta_1'-\zeta_1& \zeta_{12}' -\zeta_{12}-
\zeta_1(\zeta'_2-\zeta_2) \\
0&1&\zeta'_2-\zeta_2 \\ 0&0&1 \ea\right)\ ,
\ea \]
the proof of this statement is reduced to an application of the
factorization property (2.14)
 for $m_i=0$, $n_i=\lambda_i$ $(i=1,2)$. For a 3-point function we use the
full $SU_3$ invariance
\be \label{2e20} \ba{c}
 \left( \ba{ccc} \Lambda^1& \Lambda^2&\Lambda^3\\ Z_g^{(1)}& Z_g^{(2)}&
Z^{(3)}_g
 \ea\right)
\chi_1(b(g, \zeta^{(1)})) \chi_2(b(g, \zeta^{(2)})) \chi_3
(b(g, \zeta^{(3)})) \\
\\
= \left( \ba{ccc} \Lambda^1& \Lambda^2&\Lambda^3\\ Z^{(1)}& Z^{(2)}& Z^{(3)}
\ea\right)\ , \quad
\chi_{i}= \chi_{\Lambda^i}
\ea \ee
($Z_g$ and $b(g,\zeta)$ being given by Eqs. (\ref{2e4})-(2.7)),  without
recourse to specific properties  of the operator realization.

\noindent \underline{Proposition 2.1}. For each decomposition
\be \label{2e21}
\lambda_1^{(i)} = \sum^3_{j=1\atop j\neq i} \mu_{ij}+\nu_1\ , \quad
\lambda_2^{(i)}=
\sum^3_{j=1\atop j\neq i} \mu_{ji} +\nu_2 \ee
of the three weights  $\Lambda^{(i)}=(\lambda_1^{(i)}, \lambda_2^{(i)}),$
$i=1,2,3$, into nonnegative
integers $\mu_{ij}$ and $\nu_\alpha$, there  exists an invariant 3-point
function given by the product
\subequations
\be \label{2e22a}
\prod^3_{i,j=1\atop i\neq j} P^{\mu_{ij}}_{ij} Q^{\nu_1}_{12} Q^{\nu_2}_{21}\ ,
 \quad  P_{ij}=P(\zeta^{(i)}, \zeta^{(j)}) \quad , \ee
where $Q_{12}(Q_{21})$ are invariant 3-point functions of three (anti)quark
representations given
 by the determinant
\be \label{2e22b}
Q_{\alpha \beta} = \left\vert
\ba{ll}
\zeta^{(1)}_{\alpha \beta} -\zeta^{(2)}_{\alpha \beta}&
\zeta^{(2)}_{\alpha \beta}-
 \zeta^{(3)}_{\alpha \beta}\\
\zeta^{(1)}_{ \beta} -\zeta^{(2)}_{ \beta}& \zeta^{(2)}_{\beta}-
 \zeta^{(3)}_{\beta}\ea\right\vert \ , \quad  \alpha \neq \beta \ ,\ \
(\alpha , \beta =1, 2)\ .
\ee
\endsubequations

All polynomial 3-point invariants are linear  combinations of the expressions
(2.22);
 these expressions are linearly independent provided
\be \label{2e23}
 \nu_1\nu_2 =0 \quad . \ee
In particular, if $\lambda^{(i)}_\alpha$ are not  decomposable in the form
(\ref{2e21}), then there
 exists no invariant 3-point function.

\noindent \underline{Outline of proof}. Verifying the invariance  of Eq.(2.22)
is straightforward. The completeness proof is based on the knowledge of
$SU_3$ fusion coefficients \cite{bmw},
 which allows one to determine the number of independent 3-point invariants
as functions of $\{
\Lambda^{(i)}\}$. Here are the main steps in the argument.

The possibility to impose the restriction (\ref{2e23}) comes from the relation
\be \label{2e24}
Q_{12} Q_{21}=-P_{12} P_{31} P_{23} -P_{21} P_{13} P_{32} \quad . \ee
This is, in fact, the only relation among the 3-point blocks (2.22).
Thus, for given
 $\Lambda^{(i)}$  all expressions of the type (2.22) with
 $\mu_{ij}$ and $\nu_\alpha$ satisfying (\ref{2e21}) and (2.23)
are linearly independent.

Which of the two $\nu_\alpha$ vanishes is determined by the
 sign of the difference
\subequations
\be \label{2e25a}
\rho \equiv \frac{1}{3} (\lambda_1-\lambda_2) = \nu_1-\nu_2 \quad ,
\ee
negative $\nu_\alpha$ being excluded ; here
\be \label{2e25b}
\lambda_\alpha= \sum^3_{i=1} \lambda_\alpha^{(i)}\ ,\quad  \alpha = 1,2 \ ;\ee
\endsubequations
$\rho$ should be an integer whenever $\Lambda^{(i)}$ admit a nontrivial
3-point invariant. It follows that
\be \label{2e26}
\ell_1=\frac{1}{3}(2 \lambda_1+ \lambda_2) = \lambda_1- \rho \ ,\quad
\ell_2 =\frac{1}{3} (\lambda_1+ 2 \lambda_2) = \lambda_2 + \rho \ee
are  natural numbers in this case. Using the fact that the intermediate
weights $\mu_{ij}$ are
 nonnegative integers, we deduce that the number of independent 3-point
invariants of the type (2.22) is
\be \label{2e27}
N(\Lambda^{(1)} , \Lambda^{(2)}, \Lambda^{(3)}) = 1+{\rm min}
(\ell_1 , \ell_2)
 - {\rm max}_i (\lambda_1^{(i)} + \lambda_2^{(i)} , \ell_1 -\lambda^{(i)}_1,
 \ell_2 -\lambda_2^{(i)})  \quad . \ee
Comparison with Ref.\cite{bmw} shows that this number coincides with the
number of independent
 Clebsch-Gordan coefficients for the triple
$(\Lambda^{(1)}, \Lambda^{(2)}, \bar \Lambda^{(3)})$.

A simple example of multiplicity higher than 1 is provided by the case
in which all
$\Lambda^{(i)}$ coincide with the adjoint
 representation : $\Lambda^{(i)}= (1,1), i = 1, 2, 3$. In this case  there
are two solutions
$\mu_{ij}$ and $\bar \mu_{ij}$
of Eq.(\ref{2e21}) that respect Eq.(\ref{2e23}) (in accord with Eq.
(\ref{2e27})) :
\[ \mu_{12} =0= \mu_{23} = \mu_{31}\ , \quad  \mu_{21} = 1= \mu_{13}=
\mu_{32} \; ,\ \
\bar \mu_{ij} =1-\mu_{ij} \quad (\nu_\alpha =0) \; .\]
A natural basis of invariants in this case is provided by the ${\cal S}_3$
symmetric combination
 (\ref{2e24}) and the skew symmetric one
\be \label{2e28}
A=A(\zeta^{(1)} , \zeta^{(2)}, \zeta^{(3)})= -P_{12} P_{31} P_{23}+ P_{21}
P_{13}P_{32} \; .
\ee

We note that the current $J(z,\zeta)$, which generates the level $k$
$\widehat {su_3}$ Kac-Moody
 algebra, being a local Bose field has an overall symmetric 3-point function
that is the product of an ${\cal S}_3$
 odd function of $z_i$ with $A$ :
\be \label{2e29}
<0|J(z_1, \zeta^{(1)}) J(z_2, \zeta^{(2)}) J(z_3, \zeta^{(3)}) |0>\  =\
 \frac {kA}{z_{12} z_{23} z_{31}}\ , \quad  z_{ij}= z_i-z_j \quad . \ee

\newpage

\section{$U_q(s\ell_3)$ and its dual. A quantum homogeneous space}
We define , following Ref. \cite{frt}, $SL_q(3)$, the $q$-deformed algebra of
functions on the  group
 $SL(3, C)$, as an associative  algebra  generated by nine elements
$T^\alpha_\beta$
 ($\alpha, \beta= 1,2,3)$, including  the identity {\bf 1}, subject to the
 commutation relations :
\setcounter{equation}{0}
\be \label{3e1}
R^{\alpha \beta}_{\sigma \tau} T^\sigma_\gamma T^\tau_\delta =
T^\beta _\tau T^\alpha_\sigma
 R^{\sigma \tau}_{\gamma \delta} \; , \ee
where $R$ is the $9 \times 9$ R-matrix for the product of two ``quark''
representations
 of the quantum universal enveloping algebra
\be \label{3e2}
U_q : = U_q (s\ell_3) \; ; \ee
$T=(T^\alpha_\beta)$  is further restricted by the condition
\be \label{3e3} {\rm det}_q T =1\ , \ee
where the $q$-determinant is defined as a central element  of the algebra of
$T^\alpha_\beta$
 satisfying (\ref{3e1}). We  review the derivation of the expression for $R$
 in Subsection 1 (fixing on the way our notation and conventions for the
quantum algebra $U_q$)
 and define in Subsection 2 the main object of interest in this paper,
the $q$-deformation of the 3-dimensional homogeneous  space $SL(3,C)/B$
 of Sec. 2.1 .

\subsection{$U_q$ : definition, $R$-matrix, exchange  operator}

In this synopsis on the Hopf algebra (\ref{3e2}) we mostly follow the
notation and  conventions of Ref. \cite{it}.

 The quantum universal enveloping algebra (\ref{3e2}) has the same
Chevalley-Cartan generators
 as  $s\ell_3$ with the provision that one also uses $q^{\pm H_i} ,$ $i=1, 2,$
as elements of the algebra. The CR (2.7a) remain  unchanged,
 while Eqs.(2.7b) are replaced by :
\be \label{3e4}
[E_i, F_j] = [H_i] \delta_{ij} \quad (i,j =1,2) \ ,\quad
[X] : = \frac{q^X-\bar q^X}{q-\bar q} \quad (\bar q \equiv q^{-1}) \ee
and by the $q$-deformed Serre relations

\be \label{3e5} \ba{l}
X_i^{(2)}X_j+X_j X_i^{(2)}= X_i X_j X_i \; ,\quad X= E, F, \quad j= i \pm 1 \;
,
\\
X^{(2)}={\displaystyle \frac{1}{[2]}} X^2 \; ,\quad [2] = q + \bar q \ . \ea
\ee

The latter relations suggest the introduction of a pair of $q$-Weyl
operators $E_{12}$ and $E_{21}$ and their conjugates by
\subequations
\be  \label{3e6a}
E_{ij}=E_iE_j-\bar q E_jE_i \; ,\quad F_{ij}= F_jF_i-qF_iF_j \ee
(($ij$)=(12) or (21)), such that
\be \label{3e6b}
 E_iE_{ij}= q E_{ij} E_i \; ,\ \  F_i F_{ij}= q F_{ij}F_i \; . \ee
\endsubequations
These relations are respected (and indeed dictated ) by the coproduct
$\Delta : U_q \to U_q  \otimes U_q$ which is fixed by its values on the
generators :
\subequations
\be \label{3e7a}
\Delta(q^{\pm H_i}) = q^{\pm H_i} \otimes
 q^{\pm H_i} \quad ({\rm or } \quad \Delta (H_i) =H_i \otimes 1 +
1 \otimes H_i ) \ee
and
\be \label{3e7b}
\Delta (E_i) =E_i \otimes \bar q^{H_i} + 1 \otimes E_i\ , \quad
\Delta(F_i )= F_i \otimes
 1 + q^{H_i} \otimes F_i \ , \ee
or, alternatively,
\be \label{3e7c}
\Delta (\tilde E_i) = \tilde E_i \otimes 1+ q^{H_i} \otimes \tilde E_i\ ,
\quad  \Delta( \tilde
F_i )= \tilde F_i \otimes \bar q^{H_i} + 1 \otimes \tilde F_i \ . \ee
\endsubequations

Here, Eqs.(3.7b) and (3.7c) should be viewed as relations in the same
 quantum universal enveloping algebra with generators $X$ and
 $\tilde X$ related by

\be \label{3e8}
\tilde E_i= q^{H_i} E_i \; , \tilde F_i= F_i \bar q^{H_i} \; . \ee
We note that $\tilde E_i$ and $\tilde F_j$ satisfy the same CR
(\ref{3e4})-(3.6) as $E_i$ and $F_j$. We shall make use of both sets of
variables
 noting that $\tilde X_i$ and $-X_i$ are related by the antipode $\gamma$ :
$U_q \to U_q$ defined as an algebra antihomomorphism such that
\subequations
\be \label{3e9a}
\gamma(q^{\pm H_i}) = q^{\mp H_i} \; ,\ \
\gamma(\tilde E_i)=-E_i\; ,\ \ \gamma(\tilde F_j) =- F_j \; ,\ \
\gamma(E_i) =-E_i q^{H_i} \; ,
\ \ \mbox{etc.}\ . \ee
To  complete our review of basic notions, we need the counit
$\varepsilon$ : $U_q \to C$ which defines the trivial representation of $U_q$ :
\be \label{3e9b}
\epsilon (q^{\pm H_i})=1 \; ,\ \ \epsilon(E_i)=0=\epsilon (F_i) \; . \ee
\endsubequations
$\Delta, \gamma, \epsilon$ and the (associative) multiplication $m$ in
$U_q$ are verified to satisfy
\be \label{3e10}
m (\gamma\otimes 1)\Delta (X)=\epsilon(X)=m(1\otimes \gamma)
\Delta(X) \; . \ee

The universal $R$ matrix is defined, for  generic $q$, as an invertible
element in the
 (topological) tensor product $U_q \bar \otimes U_q$ which intertwines
the permuted coproduct
\subequations
\be \label{3e11a}
\Delta' (X)= \sigma\Delta (X) \; ,\ \  \sigma(X \otimes Y) =
Y\otimes X   \ee
with $\Delta$ :
\be \label{3e11b}
\Delta' (X)=R\Delta (X)R^{-1} \quad \mbox{for all} \; X \in U_q \; . \ee
\endsubequations
It is known by now quite explicitly \cite{r,krlskt,it} :
\subequations
\be \label{3e12a}
R= Q W_1 W_{12} W_2 \;, \ee
where
\be \label{3e12b}
Q=q^{-c_{ij}^{-1} H_i \otimes H_j} \; ,\ \  (c^{-1}_{ij}) = \frac{1}{3}
\left( \ba{ll} 2&1\\
1& 2\ea \right) \; ,\ \  Q^{-1} (1 \otimes E_i) Q= q^{H_i} \otimes E_i \; ,
\ \ {\rm etc.}\ , \ee
\be \label{3e12c}
W_\alpha =e_+(\rho E_\alpha \otimes F_\alpha) \; ,\ \  \rho =\bar q-q \; ,
\ \  \alpha= 1,2, 12\ .\ee
\endsubequations
The $q$-deformed exponents
\be\label{3e13}
 e_{\pm}(X)= \sum^\infty_{n=0} \frac{X^n}{(n)_\pm ! }\; ,\ \ (n)_\pm=
\frac{q^{\pm 2n}-1}{q^{\pm 2}-1} \; ,\ \  (n)_\pm !=(n)_\pm(n-1)_\pm ! \ee
are well defined for generic $q$ (for which $q^{\pm 2n}\neq 1)$ and satisfy
\be \label{3e14}
e_+(X)e_-(-X)=1 \; . \ee
For a finite dimensional representation of $U_q$ the series
$W_\alpha$ reduce to finite sums and make also sense for $q^p=1$
provided that the weight of the representation satisfies
$\lambda_1+ \lambda_2 <p$.

We define the {\it  exchange operator} $\hat R= \hat R^{\pi \pi' }$
 for a given pair of representations ($\pi , \pi')$ of $U_q$ as the
 corresponding $R$ followed by a permutation $P$ (acting in the tensor product
 of representation spaces):
\renewcommand{\theequation}{\arabic{section}.\arabic{equation}a}
\be \label{3e15a} \hat R=P R^{\pi \pi '}\ , \ee
where
\setcounter{equation}{14}
\renewcommand{\theequation}{\arabic{section}.\arabic{equation}b}
\be \label{3e15b}
R^{\pi \pi '} = \sum_n \pi (X_n) \otimes \pi ' (Y_n) \; \ \ \mbox{for } \;
R= \sum_n X_n  \otimes Y_n \; .
\ee

We note that the action of $U_q$ in the tensor product of $U_q$ modules is
given by
 the coproduct :

\noindent if
\renewcommand{\theequation}{\arabic{section}.\arabic{equation}a}
\be \label{3e16a}
\Delta(X)= \sum_i L_i \otimes R_i \; ,\ \  L_i =L_i (X) \; ,\ \
R_i =R_i (X) \ee
then
\setcounter{equation}{15}
\renewcommand{\theequation}{\arabic{section}.\arabic{equation}b}
\be \label{3e16b}
\Delta^{\pi \pi '}(X) = \sum_i \pi (L_i) \otimes \pi ' (R_i) \; . \ee
The exchange operator intertwines between the products $\pi \otimes \pi'$ and
 $\pi ' \otimes \pi$ :
\renewcommand{\theequation}{\arabic{section}.\arabic{equation}}
\be \label{3e17}
\hat R^{\pi \pi '} \Delta^{\pi \pi '} (X) = \Delta^{\pi' \pi} (X)
\hat R^{\pi '\pi } \; .
\ee
We shall reproduce the computation of the exchange  operator for two
 (3-dimensional) fundamental representations of $U_q$ (cf. Ref. \cite{frt}).

For $\Lambda_1 =(1,0)$ the representation $\pi _1$ of $U_q$ coincides with
the $3 \times
3$  matrix representation of the undeformed $s\ell_3$ Lie algebra ; we have  :
\subequations
\begin{eqnarray} \label{3e18}
\pi_1(E_i)=e_{ii+1}=: e_i \; ,\ \  \pi_1(F_i) = e_{i+1i} =: f_i \; ,\ \
\\
\pi_1(H_i)=e_{ii}-e_{i+1 i+1} =: h_i = [e_i , f_i]    \; .
 \end{eqnarray}
\endsubequations
 Here  $e_{ij}$ are the Weyl matrices characterized by the product formula
\be \label{3e19}
e_{ij}e_{k\ell} =\delta_{jk} e_{i\ell} \quad \quad (e_{ij} =e_{ji}^*) \;. \ee
Noting that $e^2_i=0= f_i^2$ we can write in this representation
\renewcommand{\theequation}{\arabic{section}.\arabic{equation}a}
\be \label{3e20}
R_{12}= Q_{12} +q^{1/3} \rho \sum_{k<\ell} e^1_{k \ell} e^2_{\ell k} \; , \ee
where
\setcounter{equation}{19}
\renewcommand{\theequation}{\arabic{section}.\arabic{equation}b}
\be Q_{12}= q^{1/3}  \sum_{ij} \bar q^{\delta_{ij}} e^1_{ii} e^2_{jj} \; , \ee
(the superscript on $e$ indicating the space in which it acts, so that
$e$'s with different superscripts commute). The
 notation $R_{ab}$, that involves a pair of labels (a,b) indicating
the space in a multiple tensor product
 on which  $R$ acts nontrivially, is used systematically by
Faddeev {\it et al.} \cite{frt}. The Yang-Baxter equation
 assumes in this notation the following  compact form :

\renewcommand{\theequation}{\arabic{section}.\arabic{equation}}
\be \label{3e21}
R_{12}R_{13}R_{23}= R_{23}R_{13} R_{12} \; . \ee

Here is a step in verifying Eq.(\ref{3e21}) which deciphers on the way the
operational meaning of
 the above notation. The left-hand side of Eq. (\ref{3e21}) is expressed
in terms of the Weyl generators
 as follows :
\begin{eqnarray} \label{3e22}
& & \nonumber \\
\bar q R_{12} R_{13} R_{23} =& &
{\displaystyle
\sum_{i,j,k} \bar q ^{\delta_{ij} +\delta_{jk}+ \delta_{ik}}
 e^1_{ii} e^2_{jj} e^3_{kk}
}\nonumber \\
& & {\displaystyle
+\rho \sum_{\stackrel {k < \ell}{\ i\ }} \left\{
\bar q^{\delta_{ik} + \delta_{i\ell}} (e^1_{ii} e^2_{k\ell} e^3_{\ell k} +
e^1_{k\ell } e^2_{\ell k} e^3_{ii}) \right.
}\nonumber \\
& & {\displaystyle
\left. + \bar q^{2 \delta_{ik}} e^1_{i\ell} e^2_{kk} e^3_{\ell i} \right\} +
\rho^2(
3e^1_{13} e^2_{22} e^3_{31}
}\nonumber \\
& & {\displaystyle
+ e^1_{12} e^2_{33} e^3_{21} + \bar q \sum_{k< \ell} e^1_{k \ell}
e^2_{\ell \ell} e^3_{\ell k})
 + \rho^3 e^1_{13} e^2_{22} e^3_{31}  }\nonumber \\
& &\quad \quad \quad \quad \quad \quad \quad \quad \quad \quad \quad \quad
{(\rho =\bar q-q) } \ \ .
\end{eqnarray}
(We note that Eq.(\ref{3e21}) is not verified order by order in $\rho$ ;
only the sum of the terms proportional
 to $\rho$  and $\rho^2$ cancel from both sides of the equation.)

We proceed to computing the exchange matrix $\hat R$ for a pair of
fundamental representations.

The  permutation operator in $C^3 \times C^3$ is given by :
\be \label{3e23}
P_{12}= \sum_{i,j} e^1_{ij} e^2_{ji}\ \quad (P^2_{12} =\sum_i e^1_{ii}
\sum_j e^2_{jj}  = 1_1 1_2 ) \; . \ee
It commutes with $Q_{12}$,
\be \label{3e24}
P_{12} Q_{12} =Q_{12} P_{12} =
q^{1/3} \sum_{i,j} \bar q ^{\delta_{ij}} e^1_{ij} e^2_{ji}
 \; , \ee
and symmetrizes the product of $W$'s :
\be \label{3e25}
PW_1W_{12} W_2=P + \rho \sum_{k< \ell} e_{\ell \ell} \otimes e_{kk}\ .\ee
Inserting in the expression for $\hat R$ we end up with
\be \label{3e26}
\hat R _{12} =q^{1/3} \left\{
\sum_{i,j} \bar q^{\delta_{ij}} e^1_{ij} e^2_{ji} +
\rho \sum_{k< \ell} e^1_{\ell \ell}
e^2_{kk} \right\} \;. \ee

The exchange operator thus obtained has two eigenvalues : a 6-fold
 degenerate one, $q^{-2/3}$, corresponding to a ``q-symmetric tensor''
expressed by products of  ``q-bosonic'' variables :
\be \label{3e27}
\hat R^{\alpha \beta}_{\gamma \delta} x^\gamma x^\delta =
q^{-2/3} x^\alpha x^\beta \quad {\rm or } \quad
x^\alpha x^\beta = qx^\beta x^\alpha \quad {\rm for} \quad
 \alpha < \beta \; ; \ee
and a 3-fold degenerate
 eigenvalue, $-q^{4/3}$, associated with $q$-fermions \cite{pwp,m,wzswz,hpt} :
\be \label{3e28} \ba{r}
\hat R^{\alpha \beta}_{\gamma \delta} \xi^\gamma \xi^\delta =
-q^{4/3} \xi^\alpha \xi^\beta \quad {\rm or } \quad
(\xi^\gamma )^2 =0= \xi^\alpha \xi^\beta + \bar q  \xi^\beta \xi^\alpha
  \quad {\rm for} \quad  \alpha < \beta \; . \ea \ee

\subsection{The Hopf algebra $SL_q(3)$, a Borel subalgebra and a quantum
coset space}

There are three equivalent approaches to introducing the algebra $SL_q(3)$.

(i) The Kobyzev-Manin appraoch \cite{m} introduces the generators
$T^\alpha_\beta$ of
 $GL_q(3)$ by demanding that the pair of transformations
\be \label{3e29}
x^\alpha \longrightarrow T^\alpha_\beta x^\beta \; ,\quad \xi ^\alpha
\longrightarrow T^\alpha_\beta
 \xi^\beta \ee
be nonsingular and preserve the exchange relations (\ref{3e27}) and
(\ref{3e28}), respectively.
 The $q$-determinant, defined by the relation
\be \label{3e30}
T^1_\alpha \xi^\alpha  T^2_\beta \xi^\beta T^3_\gamma \xi ^\gamma = {\rm det}_q
 (T^\alpha_\beta) \xi^1 \xi^2 \xi^3 \;, \ee
is shown to commute with $T^\alpha_\beta$ and can hence be set equal to 1.

(ii) The  Faddeev-Reshetikhin-Takhtajan approach \cite{frt} starts
with the exchange  relations (\ref{3e1}) for $T^\alpha_\beta$.

(iii) In Drinfeld's framework \cite{d} $T^\alpha_\beta$ appear as linear
functionals on $U_q$.
 Their  algebraic properties are determined  from the duality
between  the coalgebra
 structure in $U_q$ and the algebra structure in $SL_q(3)$ expressed by the
 relation
\be \label{3e31}
<A.B, X> =<A \otimes B, \Delta(X)> \quad {\rm for}\quad X \in U_q \;. \ee

We shall relate the first two approaches and will content ourselves
with a remark concerning the third one.

Assuming  that $T^\alpha_\beta$ commute with both $x^\gamma$ and $\xi^\gamma$
 we find that Eq.(\ref{3e1}) is a necessary and sufficient condition, that the
 transformed (according to Eqs.(\ref{3e29})) $x$ and $\xi$ obey the same
 exchange relations (\ref{3e27}) and (\ref{3e28}) as the original ones.
Indeed, requiring the CR
\renewcommand{\theequation}{\arabic{section}.\arabic{equation}a}
\be
\hat R_{12} T_1T_2=T_1T_2 \hat R_{12} \ee
we guarantee that $T_1T_2$ maps eigenvectors of $\hat R_{12}$
(like $x^\alpha x^\beta$ or $\xi^\alpha  \xi^\beta$) into eigenvectors
corresponding to the same eigenvalue. On the other hand,
Eq. (3.32a) is equivalent to
\setcounter{equation}{31}
\renewcommand{\theequation}{\arabic{section}.\arabic{equation}b}
\be
R_{12} T_1T_2=P_{12} T_1T_2 P_{12} R_{12} = T_2 T_1 R_{12} \ee
 which is a shorthand for Eq. (\ref{3e1}).

In exploiting Eq. (\ref{3e1}) it is convenient to write $R_{12}$ in
tensor components :
\renewcommand{\theequation}{\arabic{section}.\arabic{equation}}
\be \label{3e33}
\bar q^{1/3}R ^{\alpha \beta}_{\sigma\tau} =\bar q ^{\delta_{\alpha \beta}}
\delta^{\alpha}_\sigma
 \delta^\beta_\tau+\rho \sum_{k< \ell} \delta^\alpha_k \delta^\ell_\sigma
 \delta^\beta_\ell \delta^k_\tau \ee
$(\rho=\bar q-q)$. Inserting Eq.(\ref{3e33}) into Eq. (\ref{3e1}) we obtain
(some  of the relations below differ from those
 listed in Ref. \cite{ac}) :
\subequations
\be \label{3e34a} \ba{lll}
\bar q T^\alpha_\beta T^\alpha_\gamma &=& \bar q^{\delta_{\beta \gamma}}
T^\alpha_\gamma T^\alpha_\beta + \rho T^\alpha_\beta
 T^\alpha_\gamma \theta_{\beta>\gamma} \;, \\
& &\theta_{\beta> \gamma}=  \left\{ \ba{l}1 \quad {\rm for} \quad \beta >
\gamma \;, \\
0 \quad {\rm for} \quad \beta \le \gamma \;, \ea \right. \nonumber \\
\ea \ee
or
\be \label{3e34b}
T^\alpha_\beta T^\alpha_\gamma = q T^\alpha_\gamma T^\alpha_\beta
\quad {\rm for} \quad  \beta< \gamma\ ;
\ee
\endsubequations
\subequations
\be \label{3e35a}
\bar q T^\beta_\gamma T^\alpha_\gamma = \bar q^{\delta_{\alpha \beta }}
T^\alpha_\gamma T^\beta_\gamma + \rho T^\beta _\gamma
 T^\alpha_\gamma \theta_{\beta>\alpha} \;,\ee
or
\be \label{3e35b}
T^\alpha_\gamma T^\beta_\gamma = \bar q T^\beta_\gamma T^\alpha_\gamma
\quad {\rm for} \quad \alpha> \beta \ ;\ee
\endsubequations

\renewcommand{\theequation}{\arabic{section}.\arabic{equation}a}
\be  \label{3e36}
\bar q^{\delta_{\alpha_\beta}} T^\alpha_\alpha T^\beta_\gamma +
\rho T_\alpha^\beta T^\alpha_\gamma \theta _{\beta > \alpha } =
\bar q^{\delta_{\alpha \gamma}}
T^\beta_\gamma T^\alpha_\alpha + \rho T^\beta_\alpha
 T^\alpha_\gamma \theta_{\alpha >\gamma} \;, \ee
or, assuming $\beta \neq \alpha \neq \gamma$,
\setcounter{equation}{35}
\renewcommand{\theequation}{\arabic{section}.\arabic{equation}b}
\be
[T^\alpha_\alpha, T^\beta_\gamma] =(q- \bar q) T^\beta_\alpha T^\alpha_\gamma
(\theta_{\beta> \alpha} - \theta_{\alpha > \gamma} )
\;; \ee
for $\beta=\gamma$ Eqs. (3.36) and the antisymmetry of the commutator imply
\renewcommand{\theequation}{\arabic{section}.\arabic{equation}}
\be \label{3e37}
[T^\alpha_\beta, T^\beta_\alpha]= 0 \; , \ee
a relation that is also recovered as a special case from
\renewcommand{\theequation}{\arabic{section}.\arabic{equation}}
\be \label{3e38}
\bar q ^{\delta_{\alpha \beta}} T^\alpha_\gamma T^\beta_\alpha-
\bar q^{\delta_{\alpha \gamma}} T^\beta_\alpha T^\alpha_\gamma =
(q-\bar q) T^\beta_\gamma T^\alpha_\alpha (\theta_{\beta > \alpha} -
\theta_{\gamma > \alpha})  \; .\ee
Since for $\alpha , \beta , \gamma , \delta = 1,2,3$ at least two  indices
 always coincide the above relations (3.34)-(3.38) exhaust all possibilities.

Using Eq. (\ref{3e30}) we compute the $q$-determinant of $T$ with the result
\be \label{3e39} \ba{r}
{\rm det}_q(T^\alpha_\beta) =T^1_1T_2^2 T_3^3 + q^2 T^1_2T^2_3 T^3_1
+ q^2 T^1_3T_1^2 T_2^3-q(T^1_1T^2_3 T^3_2 \\
+ T^1_2T^2_1 T^3_3 + q^2T^1_3 T^2_2 T^3_1) \; . \ea \ee
It is verified to commute with all $T^\alpha_\beta$.

The algebra generated by $T^\alpha_\beta $ can
 be  equipped with a coproduct,
\be \label{3e40}
\Delta(T^\alpha_\beta)= \sum^3_{\sigma=1} T^\alpha_\sigma
\otimes T^\sigma_\beta \;, \ee
which respects the above commutation relations.
One can also introduce a counit $\epsilon$
 and an inverse $\gamma$, defined on the generators by :
\be \label{3e41}
 \epsilon (T^\alpha_\beta )= \delta^\alpha_\beta \;, \quad
\gamma(T^\alpha_\beta)=  (T^{-1})^\alpha_\beta \;, \ee
$T^{-1}$ being the inverse matrix of $T$, which has the form
(in view of Eqs.(\ref{3e3}) and  (\ref{3e39})):

\be \label{3e42}
T^{-1} =\left( \ba{lll}
T^2_2T^3_3-qT^2_3 T^3_2  &
T^1_3T^3_2- \bar q T^1_2 T^3_3 & \bar q^2 T^1_2 T^2_3- \bar qT^1_3 T^2_2 \\
q^2T^2_3T^3_1-q T^2_1 T^3_3 &T^1_1T^3_3- q T^1_3 T^3_1 &
T^1_3 T^2_1-\bar q T^1_1 T^2_3 \\
q^2T^2_1T^3_2-q^3 T^2_2 T^3_1 &q^2T^1_2T^3_1- q T^1_1 T^3_2 &
 T^1_1 T^2_2- q T^1_2 T^2_1
\ea \right) \ .\ee

$\Delta , \varepsilon$ and $\gamma$ along with the ordinary multiplication
$m$, can be verified
 to satisfy the defining conditions (\ref{3e10}) for a Hopf algebra.

\noindent \underline{Remark}. The elements of $SL_q$ (3) can be viewed as
 linear functionals on $U_q$, the values of which are computed as follows.

Consider the $3\times 3$ matrix representation $\pi_1$ of weight (1,0) of $U_q$
 defined by Eqs. (3.18).
 We then define the linear functional $T^\alpha_\beta$ on an arbitray
element $X$ of $U_q$ by setting
\be \label{3e43}
(T^\alpha_\beta , X) = \pi_1 (X)^\alpha_\beta \;. \ee
We then use Eq.(3.31) to extend this definition to products of $T^\alpha_\beta$
and verify that the products so obtained are noncommutative,
but satisfy Eqs.(3.32).

It is  straightforward to define the quantum Borel subalgebra $B_q$
generated by
 elements of lower triangular matrices (2.2) satisfying (in accord with
 Eqs. (3.34)-(3.38)) :

\renewcommand{\theequation}{\arabic{section}.\arabic{equation}a}
\be \label{3e44}
[\beta_1, \beta_2] = 0 \;,\ \  \beta_i b_j \beta_i^{-1} =q^{\delta_{ij}} b_j\ ,
\quad \beta_i b_{12}= q b_{12} \beta_i
\quad (i,j, =1,2) \; ; \ee
\setcounter{equation}{43}
\renewcommand{\theequation}{\arabic{section}.\arabic{equation}b}
\be
[ b_1, b_2] = (q-\bar q)  b_{12} \beta_1^{-1} \beta_2 \; ,
\ \ b_1 b_{12} =qb_{12}b_1 \; ,
\quad b_{12} b_2=qb_2 b_{12} \;.
\ee

For the subset of generators for which
\renewcommand{\theequation}{\arabic{section}.\arabic{equation}}
\be \label{3e45}
\beta_2=(T^3_3)^{-1} \quad {\rm and} \quad \beta_1=
(T^2_2 T^3_3 -q T^2_3 T^3_2)^{-1} \quad \quad
 \quad {\rm exist} \ee
the matrix ($T^\alpha_\beta)$ admits a Gauss decomposition
\be \label{3e46}
(T^\alpha_\beta) = \left( \ba{lll} 1& u_1&u_{12} \\ 0&1& u_2\\ 0&0& 1\ea
\right)
\left( \ba{lll}  \beta_1 & 0& 0 \\ b_1 & \beta^{-1}_1 \beta_2 &0 \\
 b_{12} & b_2 &\beta^{-1}_2 \ea \right)  \ee
with $\beta_1$ and $\beta_2$ given by Eqs.(\ref{3e45}) and

\be \label{3e47}  \ba{l}
b_{12} = T^3_1\ , \quad \;  b_2= T^3_2 \; , \\
b_1 \beta^{-1}_2= T^2_1 T^3_3- q T^2_3 T^3_1 \; ; \ea \ee

\be \label{3e48}
u_{12} = T^1_3 \beta_2 \;, \ \ u_2= T^2_3 \beta_2 \;,\ \
u_1=(T^1_2 T^3_3-qT^1_3 T^3_2)\beta_1 \;. \ee

By definition, $u_a$ generate the quantum coset space
\be \label{3e49} N_q =SL_q (3)/B_q \ee
(that can be viewed as an associative algebra, but not as a bialgebra).

As a consequence of Eqs. (\ref{3e48}) and (\ref{3e45}) which imply
\renewcommand{\theequation}{\arabic{section}.\arabic{equation}a}
 \be \label{3e50} \ba{ll}
[\beta_1, \beta_2] = 0 \quad \; , &\beta_2 T^i_3= qT^i_3 \beta_2
\quad \; (i=1,2) \;, \\
&\beta_1 T^1_j= qT^1_j \beta_1 \quad \; (j=1,2) \;, \ea \ee
we obtain the CR :
\setcounter{equation}{49}
\renewcommand{\theequation}{\arabic{section}.\arabic{equation}b}
 \be 
q u_1 u_2-u_2u_1=(q-\bar q) u_{12} \;, \ \ u_1u_{12}= q u_{12} u_1 \; ,\ \
u_{12}u_2=q u_2 u_{12} \; . \ee

\subsection{Quantum group action on the noncommutative coset space $N_q$}
\renewcommand{\theequation}{\arabic{section}.\arabic{equation}}
Thus, for $q^2 \neq 1$, the polynomial algebra, that defines the quantum coset
 space (\ref{3e49}), has just two generators, $u_1$ and $u_2$. Following the
 pattern of Sec. 2 we can view $N_q$ as a quantum group orbit. We shall
display the induced action of the two
 conjugate Borel subgroups $SL_q(3)$ on this noncommutative coset space.

The action of the 4-parameter q-subgroup of upper triangular matrices
is determined from
\renewcommand{\theequation}{\arabic{section}.\arabic{equation}a}
\be \label{3e51}
\left( \ba{lll}
\alpha_1& t_1\alpha_1^{-1} \alpha_2 &0 \\
0 & \alpha_1^{-1} \alpha_2 & t_2\alpha_2^{-1} \\
0& 0& \alpha_2^{-1} \ea \right)
\left( \ba{lll} 1&u_1& u_{12} \\ 0&1& u_2 \\ 0&0&1 \ea\right) =
\left( \ba{lll}
\alpha_1 & \tilde u_1\alpha_1^{-1} \alpha_2 & \tilde u_{12} \alpha_2^{-1} \\
0&\alpha_1^{-1}\alpha_2 & \tilde u_2\alpha_2^{-1}\\
0& 0& \alpha_2^{-1} \ea \right) \equiv n \; , \ee
where the right-hand side has the Gauss decomposition
\setcounter{equation}{50}
\renewcommand{\theequation}{\arabic{section}.\arabic{equation}b}
\be 
n=n(\alpha, \tilde u) = \left( \ba{lll} 1& \tilde u_1 & \tilde u_{12} \\
 0& 1 & \tilde u_2 \\  0& 0& 1 \ea \right)
\left( \ba{lll} \alpha_1 &0 & 0 \\ 0& \alpha_1^{-1} \alpha_2 &0 \\
0&0 & \alpha_2^{-1} \ea \right) \; ; \ee
here
\setcounter{equation}{50}
\renewcommand{\theequation}{\arabic{section}.\arabic{equation}c}
\be 
\ba{l} \tilde u_1= \alpha_1 u_1 \alpha_1 \alpha_2^{-1}  + t_1 \;,\ \
\tilde u_2=\alpha_1^{-1} \alpha_2 u_2 \alpha_2 +t_2\ ,\ \
\tilde u_{12} =\alpha _1 u_{12}\alpha _2 + t_1 \alpha_1^{-1}
\alpha_2 u_2\alpha_2 \; . \ea \ee

The CR among group parameters and coset space coordinates are
dictated by the requirement
 that both the first factor in Eq.(3.51a)
and the product $n(\alpha, \tilde u)$ belong to $SL_q(3)$.
 The former implies
\renewcommand{\theequation}{\arabic{section}.\arabic{equation}a}
\be \label{3e52}
\alpha_it_j= q^{\delta_{ij}} t_j \alpha_i \ ,\quad t_2t_1 = qt_1t_2 \; , \ee
while the latter gives :
\setcounter{equation}{51}
\renewcommand{\theequation}{\arabic{section}.\arabic{equation}b}
\be \alpha_iu_j =q^{\delta_{ij}} u_j\alpha_i\ , \quad
[u_i , t_j] = 0 \quad (i,j = 1,2 )\ . \ee

Similarly, lower diagonal $SL_q(3)$ transformations are deduced
from the product formula

\renewcommand{\theequation}{\arabic{section}.\arabic{equation}a}
\begin{eqnarray} \label{3e53}
& &\left( \ba{lll} \beta_1 &0 & 0 \\ b_1& \beta_1^{-1} \beta_2 &0 \\
0&b_2 & \beta_2^{-1} \ea \right) \left( \ba{lll} 1& u_1 & u_{12} \\
 0& 1 & u_2 \\  0& 0&1 \ea \right) =
\left( \ba{lll} \beta_1& \beta_1u_1& \beta_1 u_{12} \\ b_1&
b_1 u_1+\beta_1^{-1} \beta_2 & b_1u_{12}+\beta^{-1}_1\beta_2 u_2 \\
0&b_2 & \beta ^{-1}_2+b_2u_2 \ea \right) \equiv g \;,\nonumber \\
& &
\end{eqnarray}
where $g$ also admits a Gauss decomposition
\setcounter{equation}{52}
\renewcommand{\theequation}{\arabic{section}.\arabic{equation}b}
\be
g=g(b,\alpha ; v)= \left( \ba{lll} 1&v_1 &v_{12} \\ 0&1&v_2 \\ 0&0&1 \ea
\right)
\left(\ba{lll} \alpha_1 & 0&0\\ b_1& \alpha_1^{-1} \alpha_2 & 0 \\
 0& b_2 & \alpha_2^{-1} \ea \right) \ ,
\ee
with
\renewcommand{\theequation}{\arabic{section}.\arabic{equation}a}
\be \ba{ll} 
\alpha^{-1}_2= &\beta_2^{-1} +b_2 u_2 \ ,\quad  \alpha_1^{-1}=
\beta_1^{-1} +b_1 u_1 \beta_2^{-1}
 + b_1 b_2 u_{21} \;, \\
 u_{21} = & q (u_1u_2-u_{12}) = {\displaystyle \frac{qu_2u_1-u_1u_2}
{q-\bar q}} \ , \ea \ee
\setcounter{equation}{53}
\renewcommand{\theequation}{\arabic{section}.\arabic{equation}b}
\be \ba{l}
v_1= \beta_1 u_1\alpha_1 \alpha_2^{-1}\ , \quad  v_2 =( b_1u_{12}+
\beta^{-1}_1 \beta_2 u_2) \alpha_2 \;, \\
v_{12} = \beta_1 u_{12} \alpha_2 \;. \ea \ee

The requirement  that $g(b, \alpha; v)$ belongs to $SL_q(3)$
and so  does the first factor in the left-hand
 side of Eq.(3.53a) yields the relations :
\renewcommand{\theequation}{\arabic{section}.\arabic{equation}a}
\be \label{3e55}
\beta_ib_j=q^{\delta_{ij}} b_j \beta_i\ , \quad  [b_1, b_2] =0=
[\beta_1,\beta_2] \;,
\ee
\setcounter{equation}{54}
\renewcommand{\theequation}{\arabic{section}.\arabic{equation}b}
\be\ba{r}
b_i u_i=q u_ib_i\ , \quad  u_1b_2= q b_2 u_1 \;, \quad
[b_1, u_2] =0= [b_2 , u_{12} ] \;, \\
\beta_iu_j=q^{\delta_{ij}} u_j \beta_i \quad (i,j =1,2) \; .
 \ea \ee

The correct CR among the elements of the two matrices in the right-hand
side of Eq. (\ref{3e53}) is then a consequence.
It is instructive to verify, for instance, the CR
\setcounter{equation}{54}
\renewcommand{\theequation}{\arabic{section}.\arabic{equation}c}
\be
[\alpha_1^{-1} , \alpha_2^{-1} ]=0 \;,\ \
\alpha_i b_j = q^{\delta_{ij}} b_j \alpha_i\ . \ee
In establishing the first one we use the relations
\renewcommand{\theequation}{\arabic{section}.\arabic{equation}a}
\be
 u_2 u_{21} = q u_{21} u_2 \quad \;, \ee
\setcounter{equation}{55}
\renewcommand{\theequation}{\arabic{section}.\arabic{equation}b}
\be
[ b_1 u_1 \beta^{-1}_2, b_2 u_2 ] = [ \beta^{-1}_2, b_1 b_2 u_{21}]
= (\bar q-q) b_1 b_2
\beta^{-1}_2 u_{21} \;. \ee

The noncommutative version of the transformation law (2.5) for $U_q$ CSO
is defined
to map polynomials into polynomials (of the same maximal degree
$\lambda_1+ \lambda_2$ in either $u_1$ or $u_2$).

\newpage

\section{Quantum CSO. $U_q$ invariant 2-point function }
\subsection{PBW and canonical bases on the quantum coset space}
\setcounter{equation}{0}
 There are two Poincar\'e-Birkhoff-Witt  (PBW) type bases (cf. Ref. \cite{r})
in the polynomial algebra of
 $u_1$, $u_2$ :
\renewcommand{\theequation}{\arabic{section}.\arabic{equation}}
\subequations
\be
B^{(1)}_{\ell mn}=u_1^{(\ell)}(u_2u_1-\bar qu_1 u_2)^{(m)} u_2^{(n)} \;,
\ee \label{4e1a}
\be \label{4e1b}
B^{(2)}_{\ell mn}=u_2^{(n)}(u_1u_2-\bar q u_2 u_1)^{(m)} u_1^{(\ell)} \;,
\ee
\endsubequations
where we are using the shorthand notation :
\be \label{4e2}
x^{(n)} =\frac{x^n}{[n]!} \;,\ \  [n]!= [n] [n-1]! \;,\ \ [0] ! = 1\;. \ee

Clearly the middle factors are expressible in terms of the variables
$u_{ij}$ defined in
Eqs. (3.50b) and (3.54a):
\be \label{4e3}
u_1u_2-\bar q u_2u_1=(1-\bar q^2) u_{12} \; ,\ \
u_2u_1 -\bar q u_1u_2= (1-\bar q^2) u_{21} \;. \ee

The form (4.1) has the advantage of exhibiting the ``zero temperature'',
$\bar q \to 0$, limit in which the two bases are simply related to the unique
 canonical (``crystal'') basis of Lusztig \cite{l} and Kashiwara \cite{k},
defined  by :
 \be \label{4e4}
L^{(1)}_{\ell mn}= u_1^{(\ell)}u_2^{(m)}u_1^{(n)} \;,\ \
L^{(2)}_{\ell mn}= u_2^{(n)}u_1^{(m)}u_2^{(\ell)} \ \ \
 \mbox{for  } \ell + n \le  m \; . \ee
The consistency  condtion for $m= \ell +n$,
\be \label{4e5}
L^{(1)}_{\ell \ell +n n} = L^{(2)}_{\ell \ell +n n} \;, \ee
is a consequence (albeit not an easy one) of the Serre relations
\be \label{4e6}
u^{(2)}_1 u_2+u_2u_1^{(2)} =u_1u_2u_1 \;,\ \
u_1u_2^{(2)} + u_2^{(2)} u_1 =u_2u_1u_2 \; .
\ee
It follows from Luszitg's expansion formulae
\subequations
\begin{eqnarray}
L^{(1)}_{\ell mn} &=& \sum^n_{k=0} \bar q^{(m-k)(n-k)}\left[
\ba{c} \ell+n-k\\ \ell \ea \right]B^{(1)}_{\ell +n-k\ k\ m-k} \nonumber \\
&=& \sum^\ell _{k=0} \bar q^{(\ell -k)(m-k)}\left[
\ba{c} \ell+n-k\\n\ea \right]B^{(2)}_{m-k\ k\ \ell + n-k} \ ,\\
L^{(2)}_{\ell mn} &=& \sum^{\ell}_{k=0} \bar q^{(\ell-k)(m-k)}\left[
\ba{c} \ell+n-k\\n\ea \right]B^{(2)}_{\ell +n-k\ k\ m-k} \nonumber \\
&=&  \sum^n_{k=0} \bar q^{(m -k)(n-k)}\left[
\ba{c} \ell+n-k\\\ell\ea \right]B^{(1)}_{m-k\ k \ell + n-k}\ ,
\end{eqnarray}
which also demonstrate that the two PBW bases coincide in the
$\bar q\to 0$ limit . Here  we have used the $q$-binomial coefficients
\be \label{4e7c}
\left[\ba{c} m\\ \ell\ea \right] = \frac{[m]! }{[ \ell]! [m-\ell]! } \;.
\ee
\endsubequations
\indent
Preparing for the study of invariant 2-point functions we shall
introduce a variant
 of the canonical basis written in terms of one of the variables $u_i$
and the commuting
 pair ($u_{12}, u_{21})$ [Eqs. (4.3)] which allows us to exclude the
 products $u_iu_j$ :
\renewcommand{\theequation}{\arabic{section}.\arabic{equation}}
\be \label{4e8}
u_1u_2= u_{12} + \bar q u_{21} \;,\ \ u_2 u_1 =u_{21} + \bar q u_{12} \;, \ \
  \left[ u_{12} \;, u_{21} \right]  = 0 \;. \ee
The new {\it standard basis} also consists of polynomials homogeneous
with respect to
 each $u_i$ and splits into two parts depending on whether the degree in $u_1$
 exceeds the one in $u_2$ or {\it vice versa}  :
\renewcommand{\theequation}{\arabic{section}.\arabic{equation}a}
\be \label{4e9a}
S^{(1)}_{ mn \mu}  = u^{(m)}_{12} u_1^\mu u^{(n)}_{21} \;,\ \
S^{(2)}_{ mn \nu} =  u_{21}^{(n)} u_2^{\nu} u_{12}^{(m)} \;. \ee
$L^{(i)}$ and $S^{(i)}$, i=1,2, span the same subspace of polynomials and
 can be expressed in terms of each other. The counterpart of Eq. (4.5),
\setcounter{equation}{8}
\renewcommand{\theequation}{\arabic{section}.\arabic{equation}b}
\be S^{(1)}_{mn0} = S^{(2)}_{mn0} \;, \ee
is a trivial consequence of the commutativity of $u_{ij}$ [Eq. (4.8)].
 It allows us, in turn, to derive Eq. (4.5), since
\renewcommand{\theequation}{\arabic{section}.\arabic{equation}}
\begin{eqnarray}
u^m_1 u_2^{m+n} u^n_1 &=&\prod^ m_{j=1} (q^{j-1} u_{12} +   \bar q ^j u_{21} )
\prod^n_{i=1} (q^{i-1} u_{21} + \bar q^i u_{12}) \nonumber \\
 &=&u_2^n u_1^{m+n} u_2^m \;.
\end{eqnarray}
\indent
In writing down CS vectors we shall need the $U_q$ counterpart of Eq. (2.15) :
\be \label{4e11}
e_+(u_1\tilde E_1) e_+ (u_{21} \tilde E_{21} ) e_+(u_2 \tilde E_2)
= e_+(u_2\tilde E_2) e_+ (u_{12}\tilde E_{12}) e_+ (u_1 \tilde E_1) =
: E_+ (u_1, u_2) \;,
 \ee
which is an identity between the two PBW bases valid for all values of $q$.
 In the small $\bar q$ limit it reduces to the relations (\ref{4e8}).
Here we have used the
 Chevalley generators (3.8) and the counterpart of Eq. (3.6a) for $\tilde E_i$
:
\be \label{4e12}
 \tilde E_{ij}= \tilde E_i \tilde E_j-\bar q \tilde E_j \tilde E_i \ \ \
((i,j )= (1,2) \ {\rm or}\ (2,1))\ . \ee

\subsection{Quantum CS vectors. Raising operators as q-derivatives}

Equation (\ref{4e11}) suggests the following $U_q$ analogue of the CS ket
 (2.13) :
\renewcommand{\theequation}{\arabic{section}.\arabic{equation}a}
\be \label{4e13}
\Lambda(u)|0> \equiv \left (\ba{cc} \lambda_1 & \lambda_2 \\ u_1& u_2 \ea
\right ) | 0 > =
E_+ (u_1, u_2)|-\lambda_2,- \lambda_1> \; ; \ee
the conjugate bra is written as a highest weight vector times the antipode
(3.9)
of $E_+$ :
\setcounter{equation}{12}
\renewcommand{\theequation}{\arabic{section}.\arabic{equation}b}
\be
<0| \left(\ba{ll} \lambda_2 & \lambda_1\\ u_1& u_2\ea \right) =
< \lambda_1, \lambda_2|  \gamma (E_+ (u_1, u_2)) \;. \ee
We note that the order of $u_\alpha$ should not change under the map $\gamma$ :
\renewcommand{\theequation}{\arabic{section}.\arabic{equation}a}
\be \label{4e14}
\gamma \{ e_+(u_j \tilde E_j )e_+(u_{ij} \tilde E_{ij}) e_+
(u_{i} \tilde E_{i})\} =
: e_+(-u_iE_i) e_+(u_{ij} E_{ji}) e_+(-u_{j} \tilde E_{j}) : \;, \ee
where the ``normal product'' of $u$-factors is given in Eq. (4.14a) by :
\setcounter{equation}{13}
\renewcommand{\theequation}{\arabic{section}.\arabic{equation}b}
\be :u_i^\ell u^m_{ij} u^n_j : = u^n_j u^m_{ij} u^\ell_i \;. \ee

One can verify that the $U_q$ action on CS vectors agrees with the global
transformation law (3.51), (3.53) in the limit of infinitesimal
group parameters $t_i$ (or $b_i$).
 We shall display the result for the raising operators $\tilde E_i$ for which
it is particularly
 simple (and hence, useful).

The 2-parameter $SL_q(3)$ subgroup ($\alpha_1, t_1)$ acting on
$u_\alpha$ according to
Eq. (3.51c) (with $t_2=0= \alpha_2-1$) gives rise to the
 following transformation law for ordered monomials :
\[
(\alpha_1, t_1) : \ \ u_2^\ell u_{12}^m u^n_1 \to (\alpha_1^{-1} u_2)^\ell
\left ( \alpha_1 u_{12} +t_1 \alpha_1^{-1} u_2\right )^m
(\alpha_1 u_1\alpha_1+t_1)^n \alpha_1^{-\lambda _2} \;.
\]
Keeping only the linear term in $t_1$  (for $t_1 \to 0$) and moving
$t_1$ to the left, using the CR (3.52), we find:
\renewcommand{\theequation}{\arabic{section}.\arabic{equation}}
\be \label{4e15}
\delta_{t_1} u_2^\ell u_{12}^m u^n_1 =t_1 {\cal D}^+_1
u^\ell_2 u^m_{12} u^n_1 \;, \ee
where we have let $\alpha_1 \to 1$ (after commuting with $t_1$) and
introduced the $q$-derivative ${\cal D}^+_1$ satisfying :
\be \label{4e16}
{\cal D}^+_1 u_2^\ell u^m_{12} u^n_1 = q^{m-\ell}(n)_+ u^\ell_2 u^m_{12}
 u_1^{n-1} +
q^{-\ell} (m)_+ u_2^{\ell+1} u_{12}^{m-1} u^n_1 \;,
\quad (n)_{\pm } = q^{\pm (n-1)} [n] \;. \ee

A similar procedure for the subgroup ($\alpha_2, t_2)$ of
transformations (3.51c) yields :
\be \label{4e17}
(\delta_{t_2}-t_2 {\cal D}^+_2) u^\ell_2 u_{12}^m u_1^m = 0 \ee
with
\be \label{4e18}
{\cal D}^+_2 u^\ell_2 u_{12}^m u^n_1 = (\ell)_+ u^{\ell-1}_2 u_{12}^m u^n_1.
\ee
Equations (\ref{4e16}) and (\ref{4e18}) are verified if we  assume the CR
\be \label{4e19}
{\cal D}^+_i u_j = q^{c_{ij}} u_j {\cal D}^+_i +\delta_{ij} \;, \ee
where ($c_{ij})$ is the $su_3$ Cartan matrix (2.8) ; as a consequence
\be \label{4e20}
{\cal D}^+_i u_{ij} = qu_{ij} {\cal D}_i^+ +u_j \; ,\ \
{\cal D}_i^+ u_{ji} = q u_{ji}
{\cal D} ^+_i \;.
\ee
It is straightforward to verity, starting from (\ref{4e16}) and
(\ref{4e18}), that
\be \label{4e21}
(\tilde E_i- {\cal D}^+_i)\ e_+ (u_2 \tilde E_2)\ e_+ (u_{12} \tilde E_{12} )\
e_+ (u_1 \tilde E_1)\ |-\lambda_2, -\lambda_1> =0 \;. \ee
Moreover, strictly speaking, we have to rely on Eq. (\ref{4e21}) to
fix  some  remaining ambiguity
 hidden in the above derivation of Eq. (\ref{4e15}). Indeed, we have made some
arbitrary choices, which affect the overall
  power of $q$, in choosing the place of the multiplier
$\alpha_i^{ -\lambda_j}$ and in moving the
 infinitesimal variables $t_i$ to the left (rather than to the right).

We end up with a remark concerning the definition of the $q$-derivatives.

Starting from a purely algebraic set-up we are looking for an
associative algebra with four generators $u_i$,
 ${\cal D}_j$, $i,j =1,2$, subject  to the (cubic) Serre relations
(\ref{4e6}) and
\be \label{4e22}
{\cal D}^{(2)}_i {\cal D}_j+ {\cal D}_j {\cal D}_i^{(2)}=
{\cal D}_i{\cal D}_j{\cal D}_i\quad (i \neq j = 1,2) \ee
(the latter reflecting the fact that the map
$\tilde E_i \to {\cal D}^+_i$ is an
antihomomorphism of algebras) and to a quadratic relation of the
type (\ref{4e19}) (with  undetermined $c_{ij}$).

\noindent \underline {Proposition}. Under the above assumptions
$(\epsilon_i \epsilon_j c_{ij})$ is the $su_3$
  Cartan matrix for some choice of the sign factors
$\epsilon_1$ and $\epsilon_2$.

\noindent \underline {Proof}. Applying ${\cal D}_i$ to both sides of
Eq. (\ref{4e6}) we find :
\[1+q^{c_{11}} =[2] q^{c_{11}+c_{12}} \;,\quad q^{c_{12}}(1+q^{c_{11}})  =[2]
\ ,\]
\[1+q^{c_{22}} =[2] q^{c_{22}+c_{21}} \;,\quad q^{c_{21}}(1+q^{c_{22}})  =[2]
\ .\]
The proposition follows.

Our choice, ${\cal D}^+_i$ (corresponding to $\epsilon_i=1)$, is
dictated by Eq. (4.20). The $e_+$-exponent
 is an eigenfunction of ${\cal D}^+_i$ :
\be \label{4e23}
({\cal D}^+_i -a)\ e_+ (au_i) = 0 \; ; \ee
this guarantees the validity of Eq. (\ref{4e21}). It is coupled to
the expression (3.7c) of
 the coproduct  for the raising operators $\tilde E_i$ (3.8) ;
 we have the factorization property :
\be \label{4e24} \ba{l}
e_+(u_i\Delta (\tilde E_i))\ (|-\lambda_2, -\lambda_1> \otimes
|-\lambda_2' , -\lambda'_1>)\ = \\
e_+(u_i\tilde E_i)\ |-\lambda_2, -\lambda_1> \otimes e_+
(q^{-\lambda_{3-i}} u_i
 \tilde E_i) |-\lambda_2'  , -\lambda'_1> \;. \ea \ee

We note finally, that the choice (4.11) for, say, $\tilde E_{12}$
is associated with the CR
\be \label{4e25} \ba{l}
u_{12} \tilde E_{12} u_{2} \tilde E_{2}=
q^2u_2  \tilde E_{2} u_{12} \tilde E_{12} \;, \\
u_1 \tilde E_{1} u_{12} \tilde E_{12}=
q^2 u_{12}  \tilde E_{12} u_{1} \tilde E_{1} \;, \ea \ee
 which ensure the validity of the relations
\be \label{4e26} \ba{l}
e_+(u_{2} \tilde E_{2}) e_+(u_{12} \tilde E_{12})=
e_+ (u_2  \tilde E_{2}+ u_{12} \tilde E_{12}) \;, \\
e_+(u_{12} \tilde E_{12}) e_+(u_{1} \tilde E_{1})=
e_+(u_{12} \tilde E_{12}+u_{1}\tilde E_{1}) \;. \ea \ee
These relations are used in deriving a factorizing property of the
type (\ref{4e24}) for CS.

\subsection{Invaraint 2-point functions}
The invariant 2-point functions
\[ <0| \bar \Lambda (u) \Lambda (v) | 0> \; ,\ \
\Lambda=( \lambda_1, \lambda_2) \;,\ \
\bar \Lambda =(\lambda_2 , \lambda_1) \]
can be determined from the expressions (4.13)-(4.14) for CS vectors.
It is however convenient to first take into account
 the implications of $H_i$ and  $\tilde E_i $ invariances.
$H_i$ invariance implies the homogeneity condition :
\be \label{4e27}  \ba{l}
<0| \bar \Lambda (\rho^2_1 \rho_2^{-1} u_1 \;, \rho^{-1}_1 \rho^2_2 u_2)\
\Lambda (\rho^2_1 \rho_2^{-1} v_1 \;, \rho^{-1}_1 \rho^2_2 v_2 ) |0> = \\
(\rho_1 \rho_2)^{\lambda_1+ \lambda_2} <0 |\bar \Lambda (u_1,u_2)
\Lambda (v_1, v_2) |0> \;, \ \ \  \rho_i >0 \;.
\ea \ee
$\tilde E_i$ invariance yields a pair of $q$-differential equations :
\be \label{4e28}
{\cal D}_{u_i}^+ <0| \bar \Lambda (u)\Lambda(v)|0> +<0| q^{H_i}
\bar \Lambda (u) \bar q^{H_i}
 {\cal D}_{v_i}^+ \Lambda (v)|0> = 0 \;, \ee
where
\be \label{4e29}
q^{H_i}\bar \Lambda (u_i, u_j) \bar q^{H_i} = \bar q^{\lambda _i} \bar \Lambda
 (q^2u_i , \bar q u_j) \;, \ \ \ i(\neq j)= 1,2\ . \ee

The invariance conditions should be supplemented by information about the
degree of the  CS vectors with respect to each  of the variables $v_i$
(and $u_i$). These degrees can be
extracted  from Eqs. (4.13) and from the folowing properties of
the raising operators :
\be \label{4e30}
\tilde E_1^{\lambda_2+1} |- \lambda_2, -\lambda_1>=0=
\tilde E_2^{\lambda_1+1} |- \lambda_2, -\lambda_1> =
< \lambda_1 , \lambda_2 |E_i^{\lambda_i +1} \;.
\ee
Adding  to this the homogeneity condition Eq. (\ref{4e27}) we look for a
general $U_q$-invariant  2-point function of the form
\renewcommand{\theequation}{\arabic{section}.\arabic{equation}a}
\begin{eqnarray} \label{4e31}
& &<0| \bar \Lambda(u)|\Lambda(v)|0> =
\sum _{\stackrel {0\le m_i,n_i\le \lambda _i}{\ \ i=1,2\ \ \ }}
\bigg\{
A_{m_1 m_2n_1n_2} u^{(m_2)}_{12} u_1^{\nu_1} u_{21}^{(m_1)}v_{21}^{(n_2)}
v_2^{\nu_1} v_{12}^{(n_1)} \nonumber \\
& &+ B_{m_1m_2n_1n_2} u_{21}^{(m_1)} u_2^{\nu_2} u_{12}^{(m_2)} v_{12}^{(n_1)}
v_1^{\nu_2} v_{21}^{(n_2)}
+C_{m_1m_2n_1n_2} u_{21}^{(m_1)} u_{12}^{(m_2)} v_{12}^{(n_1)} v_{21}^{(n_2)}
\bigg\} \ ,
\end{eqnarray}
where $x^{(n)}$ is defined by Eq. (4.2) and
\setcounter{equation}{30}
\renewcommand{\theequation}{\arabic{section}.\arabic{equation}b}
\be
1\le \nu_i=\lambda_1+\lambda_2 -m_1-m_2-n_1-n_2 \le {\rm min}
(\lambda_i-m_i, \lambda_i - n_i) \;,\ \  i=1,2 \;, \ee
while  in the last term the sum of indices is $\lambda_1+\lambda_2$.

Invariance under $\tilde E_1$ gives the relations
\renewcommand{\theequation}{\arabic{section}.\arabic{equation}a}
\be \label{4e32}
q^\nu ([m_1] A_{m_1-1 m_2+1 n_1n_2 }+ [\nu +m_2] A_{m_1 m_2 n_1 n_2})
+ q^{\lambda_2} (A_{m_1m_2n_1+1 n_2} +
\delta_{\nu_1} C_{m_1 m_2 n_1 +1 n_2} ) =0 \;,
 \ee
\setcounter{equation}{31}
\renewcommand{\theequation}{\arabic{section}.\arabic{equation}b}
\be
q^\nu ([n_2] B_{m_1 m_2 n_1+1n_2 -1 }+ [n_1+ \nu] B_{m_1 m_2 n_1 n_2})
+ q^{\lambda_1+2} (B_{m_1m_2+1 n_1 n_2} +
\delta_{\nu_1} C_{m_1 m_2 +1 n_1 n_2} ) =0 \;,
\ee
where
\setcounter{equation}{31}
\renewcommand{\theequation}{\arabic{section}.\arabic{equation}c}
\be
\nu =\lambda_1+\lambda_2-m_1-m_2-n_1-n_2\ . \ee
$\tilde E_2$ invariance implies :
\renewcommand{\theequation}{\arabic{section}.\arabic{equation}a}
\be \label{4e33}
q^\nu([n_1]A_{m_1  m_2 n_1-1 n_2+1} +[n_2 +\nu]A_{ m_1m_2 n_1 n_2})
+q^{\lambda_2 +2} (A_{m_1+1  m_2 n_1 n_2} +
\delta_{\nu_1} C_{m_1+1  m_2 n_1 n_2} ) =0
\;, \ee
\setcounter{equation}{32}
\renewcommand{\theequation}{\arabic{section}.\arabic{equation}b}
\be
q^\nu ([m_2] B_{m_1 +1 m_2-1 n_1n_2 }+ [\nu +m_1] B_{m_1 m_2 n_1 n_2})
+ q^{\lambda_1} (B_{m_1m_2 n_1 n_2+1} +
\delta_{\nu_1} C_{m_1 m_2  n_1 n_2 +1} ) =0 \;,
 \ee
where $\nu$ is again given by Eq. (4.32c).

The basis (4.9) used above is particularly appropriate for
 $q$-symmetric tensor representations (of the type ($\lambda_1,0$) or
 ($0, \lambda_2$)). Using the normalization of (4.13) and the choice of phase
\renewcommand{\theequation}{\arabic{section}.\arabic{equation}}
\be \label{4e34}
< \lambda , 0
|\tilde E_{12}^{(\lambda)}|0,-\lambda> =q^\lambda=<0, \lambda|
 \tilde E_{21}^{(\lambda)}|-\lambda , 0> \;, \ee
we end up with :
\renewcommand{\theequation}{\arabic{section}.\arabic{equation}a}
\be \label{4e35}
<0|\left (\ba{ccc} 0& & \lambda \\ &u& \ea\right)
\left (\ba{ccc} \lambda & &0 \\ &v& \ea\right )|0> = [\lambda]!
\sum _{\stackrel {0\le m,n}{m+n\le \lambda }}\
\frac{ (-u_1v_2)^{\lambda-m-n}}{
(\lambda-m-n)_+!} (\bar q u_{21})^{(m)} (qv_{12})^{(n)} \; ,\ee
\setcounter{equation}{34}
\renewcommand{\theequation}{\arabic{section}.\arabic{equation}b}
\be
<0|\left (\ba{ccc}  \lambda & &0 \\ &u& \ea\right)
\left (\ba{ccc} 0& &\lambda \\ &v& \ea\right )|0> = [\lambda]!
\sum_{\stackrel {0\le m,n}{m+n\le \lambda }}
\frac{ (-u_2v_1)^{\lambda-m-n}}{
(\lambda-m-n)_+!} (\bar q u_{12})^{(m)} (qv_{21})^{(n)} \; .\ee
(The two expressions are obtained from one another by just
exchanging the subscripts 1 and 2.)

The structure of the invaraint 2-point function for $\lambda_1 \lambda_2>0$ is
 more  complicated. In the simplest example of this type, for the adjoint
representation $\Lambda
=(1,1)=\bar \Lambda$, the above recurrence relations and
normalization condition give :
\renewcommand{\theequation}{\arabic{section}.\arabic{equation}}
\be \label{4e36}\ba{l}
-<0|\left (\ba{ccc}  1& &1 \\ &u& \ea\right )
\left (\ba{ccc} 1& &1 \\ &v& \ea \right )|0> =
 (\bar q u_{12}+qv_{12}) (\bar q u_{21} + q v_{21})  \\
+[2] (u_{12} v_{12} + u_{21} v_{21})- u_{12} u_1v_2-
u_{21} u_2 v_1 - q^2(u_2 v_{12} v_1+u_1 v_{21} v_2)
\; . \ea\ee
In order to write down the most general 2-point invariant
we recur to a modified PBW basis looking for an expansion of the  form :
\begin{eqnarray} \label{4e37}
<0| \bar \Lambda(u)\Lambda(v)|0> =
\sum _{\stackrel {r+s=\lambda _1 +\lambda _2}{\ m,n,p\ }}
C^{rs}_{mnp}\left [\ba{c} r\\m \ea \right ]
\left [\ba{c} s\\ n \ea \right ]& &\left [\ba{c} \lambda_2\\r-p \ea \right ]
\left [\ba{c} \lambda_1\\p \ea \right ]\nonumber \\
& &\times u^p_1 u^n_{12} u^m_2 v_2^{r-m}
v_{12}^{s-n} v_1^{r-p} \;,
\end{eqnarray}
where we are using the $q$-binomial coefficients (4.7c) and the
constants $C^{rs}_{mnp}$
 will be determined from $\tilde E_i$ invariance.
The form (\ref{4e37}) is suggested
 by the classical expression (2.18)  for which the unknown coefficients
give a sign factor,
$(-1)^{m+n+r}$. In the $q$-deformed case it is consistent
to demand that each such sign is  multiplied by a power of $q$.

The condition of $\tilde E_2$ invaraince is simpler to take
into  account. Using Eq.(4.29) and the relations
\[ \ba{l}
{\cal D}^+_2 u_1^p u_{12}^n u_2^m =q^{m+n-p-1}
[m] u_1^p u_{12}^n u^{m-1}_2 \;, \\
  \\
{\cal D}^+_2 v_2^{r-m} v_{12}^{s-n} v_1^{r-p} =
 q^{r-m-1} [r-m] v_2^{r-m-1} v_{12}^{s-n} v^{r-p}_1 \;, \ea
  \]
we find
\be \label{4e38}
C^{rs}_{mnp} =-q^{r-\lambda_2-1} C_{m-1np}^{rs} =
(-1)^m q^{(r-\lambda_2-1)m} C_{0np}^{rs}
 \;. \ee
$\tilde E_1$ invariance yields a $4$-term relation; using the relations
\[ {\cal D}^+_1 u_1^p u_{12}^n u_2^m = q^{p-1} [p] u_1^{p-1} u_{12}^n u^{m}_2
+ q^{2p} [n] u_1^p u_{12}^{n-1} u_2^{m+1}
\]
and
\begin{eqnarray}
{\cal D}^+_1 v_2^{r-m} v_{12}^{s-n} v_1^{r-p} = q^{m+s-n-p-1} [r-p]& &
v_2^{r-m} v_{12}^{s-n} u^{r-p-1}_1  \nonumber \\
&+& q^{m+s-r-n-1} [s-n] v_2^{r-m+1} v_{12}^{s-n-1} v_1^{r-p}
\nonumber
\end{eqnarray}
we find :
\be \label{4e39}
\ba{l} [r] \left\{
[\lambda_1-p] C^{rs}_{mnp+1} +
q^{s-\lambda_1-1}[\lambda_2+p-r+1] C^{rs}_{mnp} \right\}\\
\\
+q^{p}[s+1]\left\{
[m]C^{r-1 s+1}_{m-1 n+1p} +q^{s+1-r-\lambda_1}[r-m]
C^{r-1s+1}_{mnp}\right\} =0 \;.
\ea \ee
Demanding that $C^{rs}_{mnp+1} =q^x C^{rs}_{mnp} $ with $x$
 chosen  in such a way that the first pair of terms in the left-hand side
is proportional to $[s+1]$ we find $x=-\lambda_1-2$.
Demanding similarly that the second pair
 of terms be proprotional to $[r]$ and  using Eq. (4.38) and the previous
result we find $C_{mn+1p}^{rs} = -q^{-1} C_{mnp}^{rs}$. Imposing finally
Eq. (4.39) we end up with the phase factors :
\be \label{4e40}
C^{rs}_{mnp}= (-1)^{r+m+n} q^{r(\lambda_1+(3-r)/2)+
\lambda_1-n-(\lambda_1+2)p-(\lambda_2+1-r)m} \quad \quad
(s=\lambda_1+\lambda_2-r)\ . \ee
It is a straightforward exercice to verify, using
\be \label{4e41}
u_{21}^{(n)}= \sum_{\ell =0}^n q^n\frac{1}{(\ell)_{-}!} u^\ell_1
(-u_{12})^{(n-\ell)} u_2^\ell \;, \ee
that the 2-point functions (4.35) and (4.36)  are then reproduced as
special cases of  Eqs.(\ref{4e37}) and (\ref{4e40}).

\vspace{1 cm}
\bc {\large Acknowledgments} \ec

I. Todorov enjoyed the hospitality of l'Institut de Physique Nucl\'eaire
in 1992  and of the
 Laboratoire de Physique Th\'eorique et Hautes Energies in November
  1993 -both at Universit\'e de Paris XI, Orsay- during the course
 of this work.
Ya. Stanev thanks Laboratorio Interdisciplinare per le Scienze Naturali
ed Umanistiche of SISSA for hospitality.
Ya. Stanev and I. Todorov also acknowledge partial support
by the Bulgarian Science Foundation for Scientific Research
under contract F-11.

\end{document}